\documentclass[twocolumn,showpacs,preprintnumbers,amsmath,amssymb]{revtex4}

\usepackage{graphicx}
\usepackage{dcolumn}
\usepackage{bm}

\usepackage{color}

\usepackage{mathptmx}
\usepackage{latexsym}
\usepackage{keyval}
\usepackage{fancybox}

\newcommand{\be}{\begin{equation}}
\newcommand{\ee}{\end{equation}}
\newcommand{\bee}{\begin{eqnarray}}
\newcommand{\eee}{\end{eqnarray}}

\newcommand{\FELA}{F^{ela}} 	
\newcommand{\FVIS}{F^{vis}}	
\newcommand{\FADH}{F^{adh}}	
\newcommand{\FEXT}{F^{ext}}
\newcommand{\XDot}{\dot X}
\newcommand{\hDot}{\dot h}
\newcommand{\dDot}{\dot \delta}
\newcommand{\aDot}{\dot a}
\newcommand{\ad}{a_\delta}
\newcommand{\ah}{a_h}

\newcommand{\FVIShd}{\zeta}

\newcommand{\FEXTDot}{\dot F^{ext}}

\newcommand{\FELADot}{\dot F^{ela}}
\newcommand{\FELAd}{\FELA_{\delta}}
\newcommand{\FELAa}{\FELA_{a}}

\newcommand{\bL}{\begin{Large}}
\newcommand{\eL}{\end{Large}}

\begin{document}

\preprint{}

\title{\textit{Soft Dynamics} simulation:\\ 
normal approach of deformable particles in a viscous fluid}

\author{Pierre Rognon}
\author{Cyprien Gay}

\email{cyprien.gay@univ-paris-diderot.fr}

\affiliation{%
Centre de Recherche Paul Pascal, CNRS UPR~8641 - Av Dr Schweitzer, Pessac, France\\
Mati\`{e}re et Syst\`{e}mes Complexes, Universit\'{e} Paris-Diderot - Paris 7, CNRS UMR~7057 - Paris, France
}%

\date{\today}

\begin{abstract}

Discrete simulation methods are efficient tools to investigate the complex behaviors of complex fluids made of either dry granular materials or dilute suspensions.
By contrast, materials made of soft and/or concentrated units (emulsions, foams, vesicles, dense suspensions) can exhibit both significant elastic particle deflections (Hertz-like response) and strong viscous forces (squeezed liquid). 
We point out that the gap between two particles is then not determined solely by the positions of their centers, but rather exhibits its own dynamics. 
We provide a new discrete numerical method, named \textit{Soft Dynamics}, to simulate the combined dynamics of particles and contacts. 
As an illustration, we present the results for the approach of two particles. 
We recover the scaling behaviors expected in three limits: the \textit{Stokes} limit for very large gaps, the \textit{Poiseuille-lubricated} limit for small gaps and even smaller surface deflections, and the \textit{Hertz} limit for significant surface deflections.
Larger scale simulations with this new method are a promising tool for investigating the collective behaviors of many complex materials.
\end{abstract}

\maketitle

Foams, emulsions and granular materials are made of interacting particles, 
respectively bubbles, droplets and grains, in a surrounding fluid. A great deal of research revealed their elastic, plastic and viscous characters~\cite{Weaire01,Coussot05}. How are these behaviors related to the particle interactions? This issue is central in the current debate about constitutive laws, dilatancy~\cite{Weaire03,Rioual05}, and shear banding~\cite{Debregeas01,Kabla03,Huang05,Becu05,Becu07}.
Discrete simulation methods are efficient tools for such an investigation, since the motion of each particle is derived from the interactions directly. Molecular Dynamics (MD) for dry granular~\cite{Cundall79}, 
Stokesian Dynamics (SD) for viscous suspensions~\cite{Durlofsky87}
and Bubble Model (BM) for foams~\cite{Durian95}
implement various adjustable particle interactions
and provide the corresponding macroscopic behaviors, with good agreement with experiments~\cite{GDR04,Brady01,Hohler05}. Hence, implementing relevant interaction models is crucial. Accurate models are already available for either deformable elastic-like grains without surrounding fluid (MD for dry granular), or non-deformable particles in a viscous fluid  (SD for suspension). 
By contrast, the combination of both particle deformation and viscous flow has not been fully described yet,
although it is central in such materials as foams and emulsions.
BM provided a first step in that direction. Like with MD for grains, BM particles can slightly overlap with an associated normal elastic-like repulsion. Besides, a tangential viscous force arises from the shearing of the contact film. However, BM does not include any viscous effects for normal displacement, and is therefore not fully suitable for the normal approach or separation of particles, which are essential in T1 and other topological processes. 

\begin{figure}[!htb]
\begin{center}
\resizebox{0.7\columnwidth}{!}{%
\input{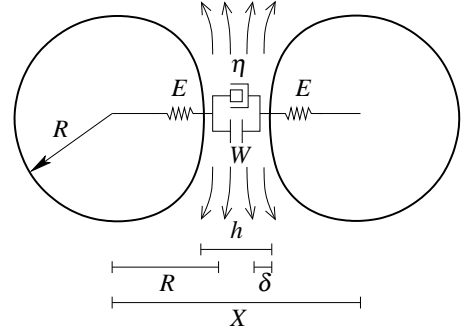}
}
\end{center}
\caption{Sketch of normal particle interaction. A force between elastic surfaces (modulus $E$) is transmitted partly through the fluid (visosity $\eta$) and partly through possible remote interaction (labeled $W$). It can deform them (deflection $\delta$), which affects the flow. The consequence is that the center-to-center distance $X$ and the gap $h$ between both surfaces exhibit two distinct dynamics.
}
\label{Fig1}
\end{figure}

In this paper, we address the description of normal interactions between two elastic-like particles in a viscous fluid, as depicted in Fig.~\ref{Fig1}. We
show that such a system exhibits two distinct dynamics for the center-to-center and surface-to-surface distances.
As this feature was ignored so far in existing simulations,
we here introduce a new discrete numerical method, named \textit{Soft Dynamics},
and implement the simple case of two similar particles for illustration purposes. 
We recover the scaling behaviors expected 
in Stokes, Poiseuille-lubricated and Hertz limits.

\textit{Model of interaction}. Let us describe a minimal model of interaction between two identical, elastic spheres 
with Young modulus $E$ and radius $R$ within a Newtonian fluid of viscosity $\eta$. 
We neglect both particle and fluid inertia, thereby assuming the system to be in a Stokes regime. Furthermore, we focus on normal motion. An external force $\FEXT$  induces a relative motion of particles, so that both their inter-centre distance $X$ and the gap $h$ evolve in time. Particle surfaces are then submitted to the viscous force $\FVIS$ due to the motion of the fluid in the gap, and also to a possible remote interaction $\FADH$ that is not transmitted through the fluid. Submitted to these two forces, particle deform elastically (surface normal deflection $\delta$, counted as positive while in compression). The force balance (with no inertia) and the geometry yield the following equations of the system (Fig.~\ref{Fig1}):

\bee \label{Eqn:equal_force}
\FEXT &=& \FELA  = \FVIS  + \FADH , \\ 
X &=& 2R+h-2\delta . \label{Eqn:DISTANCE}
\eee

\noindent According to Hertz~\cite{Hertz82}, the elastic deformation of a sphere due to a force acting on a region of radius $a$ is mainly located within a volume of size $a$, and scales like $\delta/a$. Hence, the total elastic energy stored in the contact scales like
$Ea\delta^2$ and the associated force reads: 

\begin{equation}\label{Eqn:FELA}
\FELA (a,\delta) = \alpha E a \delta.
\end{equation}

\noindent where the geometrical constant $\alpha$ is equal to $4/9$ for elastic solids.
The Young modulus $E$ ranges typically from 100 GPa for metals 
to 100 MPa for polymers or 1 MPa for biomaterials. 
By contrast, Eq.~(\ref{Eqn:FELA}) is not so accurate for bubbles and droplets 
since their bulk is not elastic. 
Nevertheless, due to the surface tension, $\sigma$, 
their surface deflects in an elastic-like manner. 
The effective Young modulus is much smaller than for solids 
since it scales like $\sigma/R$ and thus ranges 
from $1$ $Pa$ to $10^4$ $Pa$ 
($\sigma \approx 10^{-2}\,{\rm Nm}^{-1}$, 
$R \approx 1\,{\rm cm}$ to $1\,\mu{\rm m}$). 
Although other parameters can play a role 
in the elastic-like properties of bubbles and droplets, 
such as the number of contacts~\cite{Lacasse96a,Lacasse96b}, 
the elastic-like force mainly depends on the deflection $\delta$ 
and size $a$ of the interacting region.

\begin{figure}[!htb] 
\begin{center}
\resizebox{.7\columnwidth}{!}{%
\input{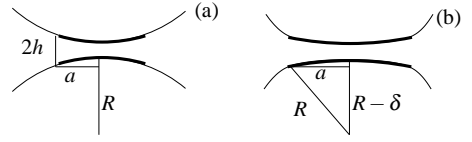}
}
\end{center}
\caption{\label{Fig:hertz_Pois} Scheme of the region through which both particles interact (bold lines). (a) In the \textit{Poiseuille lubricated} limit for sligthly deflected surfaces ($\delta \ll h$), the size $a$ of the interacting region depends mainly on $h$ and marginally on $\delta$. (b) By Contrast, in the \textit{Hertz} limit for significant deflections ($\delta \gg h$), it depends mainly on $\delta$ and marginally on $h$.}
\end{figure}

The viscous force exerted by the fluid on the particle surfaces, 
due to their normal motion $\hDot$, is well established in three limits. 
First, the \textit{Stokes} limit holds for very large gaps ($h \gg R$):
the viscous force $6\pi \eta R \hDot$ is transmitted 
through a region of size $a \approx R$. 
Secondly, for smaller gaps ($h \lesssim R$), 
and negligible surface deflections ($\vert \delta \vert\ll h$), 
the interaction can be regarded as a \textit{Poiseuille lubrication} flow 
between two spheres. 
Hence, the viscous force $\frac{6\pi \eta R^2}{h} \hDot$ 
is mainly transmitted through a region of size $a$ 
where the surface-to-surface distance lies between $h$ and $2h$, 
yielding $a \approx \sqrt{2Rh}$ (Fig.~\ref{Fig:hertz_Pois} a). 
Thirdly, the \textit{Hertz} limit corresponds to elastic deflections 
larger than the gap ($\vert \delta \vert \gg h$). 
In this case, the force is transmitted mainly through a region of size $a$ 
determined by Hertz theory, $a \approx \sqrt{2R \vert \delta \vert}$, 
and the contact can be regarded as two parallel disks, 
so that the viscous force reads $\frac{ 3\pi \eta a^4 }{ 2h^3} \hDot$. 
In order to interpolate between these three limits, 
we express the viscous force as:

\begin{equation}\label{Eqn:FVIS}
\FVIS (a,h,\hDot)= -\FVIShd \hDot,
\end{equation}

\noindent where $\FVIShd(a,h) = \frac{ 3\pi \eta a^4 }{ 2\overline{h}^3}$ 
is the effective normal friction coefficient of the contact,
with $\overline{h} = Rh/(R+h)$. 
Other choices are possible for this coefficient, but it will at least depend on the contact geometry through $h$ and $a$. We choose an expression for the size of the interacting region that interpolates between the asymptotic expressions valid in all three regimes (Stokes, Poiseuille and Hertz):

\be
\label{Eqn:surface_contact}
a(h,\delta) = \sqrt{2R (\overline{h}+\vert \delta \vert)}
\ee

\noindent Other choices for $a$ are possible, but $a$ will depend at least on $h$ and $\delta$.

A remote interaction between both particle surfaces can also be present. It may include steric repulsion (with a closest approach distance related to the size of the surface asperities in solid grains), surfactant repulsion or, more generally, disjoining pressure in foams or emulsions~\cite{Stubenrauch04}. It may also include some adhesion. Various models can be developped for such remote forces, which will at least depend on the gap $h$ and on the size $a$ of the interacting region: $\FADH (a,h)$. In the present work, for simplicity, we omit such an interaction.

\begin{figure*}[!htb]
\begin{center}
\resizebox{2.\columnwidth}{!}{%
\input{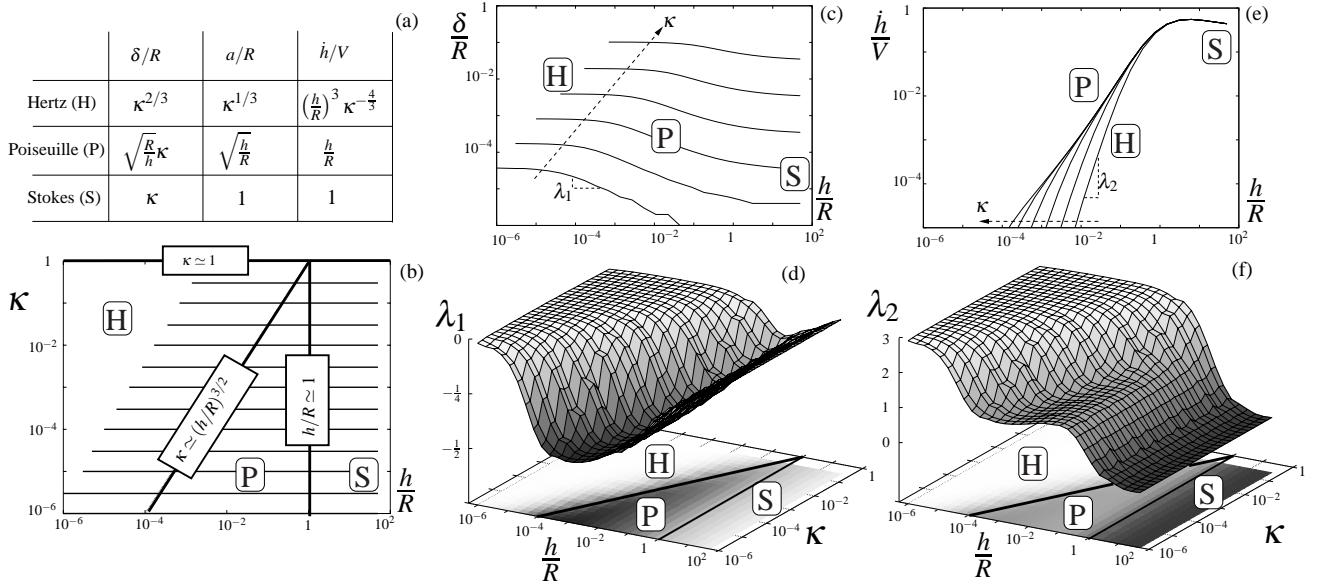}
}
\end{center}
\caption{
Scaling behaviors of two elastic-like sphere in a viscous fluid. (a) Scaling expected from theory in the Stokes, the Poiseuille-lubricated and the Hertz limits; the elastic deflection $\delta /R$, the contact size $a/R$ and the gap rate of change $\hDot /V$ are function of the current gap $h/R$ and renoralized force $\kappa$. (b) Phase-diagram expected from theory for the crossover (large lines) between the three limits ; lines at constant $\kappa$ come from Soft Dynamics simulations (see Fig.~\ref{Fig:methode_numerique}). (c-f) $\delta (h)$, $\hDot (h)$ and their power law also from these simulations. Theoretical phase-diagram is replotted on the 3D plot basis.}
\label{Fig:comparison}
\end{figure*}

Let us now consider two particles with elastic deflection and viscous interactions.
This situation evokes a Mawxell fluid whose elastic and viscous coefficients would depend on the gap $h$ and on the size $a$ of the interacting region. 
We can derive the expected scaling behaviors of $\delta$, $a$ and $\hDot$ with respect to $h$ in several asymptotic regimes, and express them simply (see Fig.~\ref{Fig:comparison} a,b) in terms of a typical Stokes velocity $V$ and a renormalized applied force $\kappa$ defined by:

\be
V = \FEXT / (\eta R), \label{Eq:stokes_velocity_kappa}
\hspace{0.8cm}
\kappa = \FEXT / (ER^2).
\ee

\noindent We do not address large elastic deformation ($\delta \simeq R$) that corresponds to  $\kappa >1$. 
In the Stokes regime ($h \gg R$) the deflection $\delta$ can be neglected in a first approximation, and we find $\hDot\simeq V$ and $a \simeq R$. According to Eq.~(\ref{Eqn:FELA}), we then obtain the deflection $\delta \simeq \kappa$. The crossover to the Poiseuille-lubricated regime occurs when the gap $h$ becomes comparable to $R$, while remaining much larger than $\delta$.
Again, in the Poiseuille regime, $\delta$ can be neglected at first, and we have $a\simeq\sqrt{Rh}$. Thus, from Eqs.~(\ref{Eqn:FVIS}) and~(\ref{Eq:stokes_velocity_kappa}), $\hDot \simeq V h/R$. Then, from Eqs.~(\ref{Eqn:FELA}) and~(\ref{Eq:stokes_velocity_kappa}), we get $\delta \simeq \sqrt{(R^3/h)} \kappa$. 
The crossover between Poiseuille and Hertz regimes occurs when the elastic deflection becomes comparable to the gap $\delta \simeq h$,
which corresponds to $\kappa \simeq (h/R)^{3/2}$. 
In the Hertz regime, $a \simeq \sqrt{R\delta}$. According to Eqs.~(\ref{Eqn:FELA}) and~(\ref{Eq:stokes_velocity_kappa}), 
this implies $a \simeq R \kappa^{1/3}$. Finally, using Eqs.~(\ref{Eqn:FVIS}) and~(\ref{Eq:stokes_velocity_kappa}), we obtain $\hDot \simeq V \kappa^{-4/3} (h/R)^3$.

\textit{Soft Dynamics simulation}. Let us now show that the gap $h$ is not only dictated by the center-to-center position $X$, but exhibits its own dynamics. The system evolution is determined by Eqs.~(\ref{Eqn:equal_force}) and~(\ref{Eqn:DISTANCE}), and by the expressions for the forces $\FELA (a,\delta)$, $\FADH (a,h)$, $\FVIS (a,h,\hDot)$ and $a(h,\delta)$. The viscous force (Eq.~\ref{Eqn:FVIS}) and force balance (Eq.~\ref{Eqn:equal_force}) directly yield the gap evolution:
\be \label{Eqn:h_dynamics}
\hDot = -\frac{\FEXT-\FADH }{\FVIShd}.
\ee
\noindent This is a purely viscous behavior with an effective friction $\FVIShd(a,h)$ and an applied force  $\FEXT-\FADH$ which corresponds to the part of the total force that is transmitted through the fluid. 
Obtaining the dynamics of $X$ is also straightforward, although it requires more steps. 
To this aim, it proves helpful to express time derivative of $a$ and $\FELA$ in terms of their partial derivatives (denoted by a subscript, for example $\FELAa = \partial\FELA/\partial a$):

\be
\aDot = \ah \hDot + \ad \dDot, 
\hspace{0.8cm}
\FELADot = \FELAa \aDot + \FELAd \dDot.
\ee

\noindent Substituting $\aDot$ and $2\dDot = \hDot - \XDot$ 
(from Eq.~\ref{Eqn:DISTANCE}) into the expression of $\FELADot$ 
yields 
$\FELADot = - {\cal C} \XDot + \hDot ({\cal C}+\FELAa \ah)$, 
where ${\cal C} = (\FELAa \ad+ \FELAd)/2$. 
Then, using the force balance $\FELADot = \FEXTDot$ 
and the expression for $\hDot$ (Eq.~\ref{Eqn:h_dynamics}) 
leads to the dynamics equation for $X$:

\be \label{Eqn:X_dynamics}
\XDot = - \frac{\FEXTDot}{{\cal C}} - \frac{ \FEXT-\FADH}{{\cal D}},
\ee

\noindent with ${\cal D} = {\cal C}\FVIShd/({\cal C}+\FELAa \ah)$. 
The above expression evokes a Maxwell behavior, involving the time derivative of the external force, 
with stiffness coefficient ${\cal C}$ and friction coefficient ${\cal D}$. 
It thus differs from the purely viscous behavior of the gap by itself 
(Eq.~\ref{Eqn:h_dynamics}). 
Both dynamics of $X$ and $h$ are here expressed in a generic and explicit manner 
in terms of the partial derivatives $\FELA_{a,\delta}$ and $a_{h,\delta}$
and the effective friction coefficient $\FVIShd$,
and also a possible remote force $\FADH$.
The expressions for $\FELA(a,\delta)$, $\FADH(a,h)$ and $a(h,\delta)$
can therefore be changed at will to describe 
various physical contact behaviours. 
The only restriction is that the viscous force 
must be linear in $\hDot$, consistently with the no-inertia assumption.

\begin{figure}[!tb]
\begin{center}
\resizebox{.9\columnwidth}{!}{%
\input{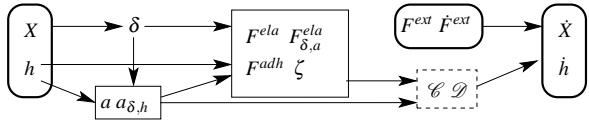}
}
\end{center}
\caption{Method for the \textit{Soft Dynamics} simulation. 
From current center-to-center distance $X$ and gap $h$, 
their respective time evolutions are explicitely calculated 
using the value of the external force $\FEXT$ and its time derivative.
The intermediate calculations involve analytical expressions 
that result from the physical description of the system 
(solid rectangular boxes).
The dashed rectangular box (with ${\cal C}$ and ${\cal D}$) 
represents intermediate calculation steps.}
\label{Fig:methode_numerique}
\end{figure}

Knowing the distinct dynamics of $h$ and $X$, 
we can now propose a method to simulate their time evolution 
in the presence of a known external force $\FEXT(t)$
(see Fig.~\ref{Fig:methode_numerique}). 
Initial values of $X$ and $h$ yield the deflection $\delta$ 
according to Eq.~(\ref{Eqn:DISTANCE}), 
and then the size of the contact $a(h,\delta)$. 
Quantities $\FVIShd$, ${\cal C}$ and ${\cal D}$ 
are explicitely given by the expression of the forces and $a$. 
Finally, the external force $\FEXT$ and its time derivative 
are needed to obtain both $\hDot$ and $\XDot$. 
By integration, new values for $h$ and $X$ are obtained.

We implemented the Soft Dynamics method 
in the simple case of two identical particles.
They are initially separated by a large distance ($h/R = 100$),
we then subject them to a constant, compressive force $\FEXT$, 
and the gap decreases continually.
$h$ may even tend to zero asymptotically since
in this case, for simplicity, we have not included the remote force $\FADH$. 
For various applied forces ($10^{-6}<\kappa<0.3$), 
Fig.~\ref{Fig:comparison}b 
shows the range of gaps $h$ reached throughout the approach. 
Plotting $\delta$ and $\hDot$ as a function of the current value of the gap $h$ 
allows us to extract the corresponding scaling behaviors. 
Figs.~\ref{Fig:comparison}c,d represent function $\delta(h)$ 
and the associated exponent $\lambda_1(h)$ 
such that one locally has $\delta(h) \propto  h^{\lambda_1}$. 
In both the Hertz and the Stokes regimes, the deflection $\delta$ 
depends only on the contact force, 
with the usual $2/3$-power law in the Hertz regime 
and a linear law in the Stokes regime 
where the interacting region spans over one entire side of the particle. 
In the Poiseuille-lubricated regime, 
the width of the interacting region depends on the gap $h$:
it causes $\delta$ to vary like $(h/R)^{-1/2}$.
Function $\hDot(h)$ is represented on Figs.~\ref{Fig:comparison}e,f 
together with the associated local exponent: $\hDot(h) \propto  h^{\lambda_2}$. 
In the Stokes regime, the large gap $h\gg R$ implies that the interaction, 
mediated by the surrounding liquid, does not depend on distance: 
$\hDot$ is then independent of $h$. 
By contrast, in both the Poiseuille and the Hertz regimes, 
the gap thickness affects the liquid flow, 
and $\hDot$ varies respectively 
like $\hDot\propto h^{1}$ and $\hDot\propto h^{3}$. 
Note that the range of gaps $h$ 
where $\lambda_1 \approx -1/2$ and $\lambda_2 \approx 1$
shrinks and vanishes as the renormalized force $\kappa$ is increased:
strongly compressed particles 
cross over from Stokes to Hertz regimes directly as $h$ deacreases.
All these observations are consistent with the theoretical predictions 
summarized on Fig.~\ref{Fig:comparison}a.

The conclusion of this work is that particles interacting 
in a viscous fluid can exhibit two distinct dynamics,
which govern their overall motion on one hand
and the gap evolution on the other hand. 
This behavior results from the interplay between the underlying 
Poiseuille and Hertz mechanisms: 
the viscous flow in the gap affects the transmitted force
and hence the surface deflection,
which in turn affects the flow in the gap.
In Stokes and Poiseuille regimes, grain surfaces are remote 
and the deflection can be neglected:
the gap dynamics is mainly dictated by the motions of the particle centers.
However, as soon as the deflection becomes larger than the gap,
both dynamics must be distinguished.
This is easily encountered with soft particles,
but can also occur with rather hard particles
as soon as the distance $h$ becomes small enough.
It seems that this feature had not been implemented so far:
we here provided a new simulation method, called \textit{Soft Dynamics}, 
which expresses each of these two dynamics generically 
in terms of the forces involved (elastic, viscous 
as well as possible remote interaction).
Here, we merely check that it describes the normal approach 
of two particles in a viscous fluid correctly.
But Soft Dynamics can be implemented
for many multicontacting particles%
~\footnote{We briefly depict here how does the ``Soft Dynamics'' approach
works with several multi-contacting particles. 
Eqs.~(\ref{Eqn:h_dynamics}) and~(\ref{Eqn:X_dynamics})
now apply to each pair $ij$ of contacting particles,
with center-to-center vectors $X_{ij}=X_j-X_i$,
gaps $h_{ij}$, 
and coefficients that vary from pair to pair.
For each particle $i$, let us use the force balance equation
$\FEXT_i=\sum_j\;\FELA_{ij}$ (see Eq.~\ref{Eqn:equal_force})
and sum up the new Eqs.~(\ref{Eqn:X_dynamics}) 
that correspond to neighbouring particles $j$.
We thus obtain the following system of vector equations
(one equation per particle):
$$
\sum_j{\cal C}_{ij}\;\XDot_j 
- \XDot_i \sum_j\;{\cal C}_{ij}  
= -\FEXTDot_i
+\sum_j \frac{{\cal C}_{ij}}{{\cal D}_{ij}}\;
\left( \FELA_{ij}-\FADH_{ij} \right) 
\label{Eq:systemXDot}
$$
Inverting this system of linear equations
yields the particle center velocities $\XDot_i$
(in fact, these equations are not independent:
one of them must be replaced, for instance, 
by the condition that the average
particle velocity is zero).
Using the Eqs.~(\ref{Eqn:h_dynamics})
then provides the gap evolutions.
Thus, when several particles are present,
the very last step of the method 
pictured on Fig.~\ref{Fig:methode_numerique}
now involves the inversion of a system of equations.}.
Hence, like MD for dry grains and SD for diluted suspensions, 
Soft Dynamics should be able to provide new insights 
into the behavior of materials made of soft, close-packed units. 
Bubbly liquids, vesicules, foams, emulsions, micellar solutions and dense suspensions
could thus be investigated,
by including the specificities of their interactions. 


We gratefully acknowledge fruitful discussions with Fran\c{c}ois Molino
and with participants of the GDR 2352 Mousses (CNRS).
This work was supported by the Agence Nationale de la Recherche (ANR05).




\end{document}